\documentstyle[12pt]{article}

\catcode`\@=11

\newcount\hour
\newcount\minute
\newtoks\amorpm \hour=\time\divide\hour by 60\minute
=\time{\multiply\hour by 60 \global\advance\minute by-\hour}
\edef\standardtime{{\ifnum\hour<12 \global\amorpm={am}%
        \else\global\amorpm={pm}\advance\hour by-12 \fi
        \ifnum\hour=0 \hour=12 \fi
        \number\hour:\ifnum\minute<10
        0\fi\number\minute\the\amorpm}}
\edef\militarytime{\number\hour:\ifnum\minute<10
0\fi\number\minute}

\def\draftlabel#1{{\@bsphack\if@filesw {\let\thepage\relax
   \xdef\@gtempa{\write\@auxout{\string
      \newlabel{#1}{{\@currentlabel}{\thepage}}}}}\@gtempa
   \if@nobreak \ifvmode\nobreak\fi\fi\fi\@esphack}
        \gdef\@eqnlabel{#1}}
\def\@eqnlabel{}
\def\@vacuum{}
\def\marginnote#1{}
\def\draftmarginnote#1{\marginpar{\raggedright\scriptsize\tt#1}}
\def\draft{
        \pagestyle{plain}
        \overfullrule=2pt
        \oddsidemargin -.5truein
        \def\@oddhead{\sl \phantom{\today\quad\militarytime} \hfil
        \smash{\Large\sl DRAFT} \hfil \today\quad\militarytime}
        \let\@evenhead\@oddhead
        \let\label=\draftlabel
        \let\marginnote=\draftmarginnote
        \def\ps@empty{\let\@mkboth\@gobbletwo
        \def\@oddfoot{\hfil \smash{\Large\sl DRAFT} \hfil}
        \let\@evenfoot\@oddhead}
        \def\@eqnnum{(\theequation)\rlap{\kern\marginparsep\tt\@eqnlabel}%
        \global\let\@eqnlabel\@vacuum}  }

\newcommand{\rf}[1]{(\ref{#1})}
\renewcommand{\theequation}{\thesection.\arabic{equation}}
\renewcommand{\thefootnote}{\fnsymbol{footnote}}

\def\appendix#1{\addtocounter{section}{1}\setcounter{equation}{0}
\renewcommand{\thesection}{\Alph{section}}
\section*{Appendix \thesection\protect\indent \parbox[t]{11.15cm}{#1}}
\addcontentsline{toc}{section}{Appendix \thesection\ \ \ #1}}

\def\nline{\,\nabla\kern -0.7em\raise0.2ex\hbox{/}\,\,}
\def\yline{\,y\kern -0.47em /}
\def\aline{\,a\kern -0.49em /}
\def\parline{\,\partial\kern -0.55em /\,\,}

\def\be{\begin{equation}}
\def\ee{\end{equation}}

\begin{document}


\begin{flushright}
FIAN/TD/19-03 \\
hep-th/0312297
\end{flushright}

\vspace{1cm}

\begin{center}

{\Large \bf Massive totally symmetric fields in AdS(d)}

\vspace{2.5cm}

R.R. Metsaev\footnote{ E-mail: metsaev@lpi.ru }

\vspace{1cm}

{\it Department of Theoretical Physics, P.N. Lebedev Physical
Institute, \\ Leninsky prospect 53,  Moscow 119991, Russia }

\vspace{3.5cm}

{\bf Abstract}

\end{center}

Free totally symmetric arbitrary spin massive bosonic and
fermionic fields propagating in AdS(d) are investigated. Using the
light cone formulation of relativistic dynamics we study bosonic
and fermionic fields on an equal footing. Light-cone gauge actions
for such fields are constructed. Interrelation between the lowest
eigenvalue of the energy operator and standard mass parameter for
arbitrary type of symmetry massive field is derived.

\newpage
\setcounter{page}{1}
\renewcommand{\thefootnote}{\arabic{footnote}}
\setcounter{footnote}{0}

\section{Introduction}

A study of higher spin theories in $AdS$ space-time has two main
motivations (see e.g. \cite{vas1,met8}): Firstly to overcome the
well-known barrier of $N\leq 8$ in $d=4$ supergravity models and,
secondly, to investigate if there is a most symmetric phase of
superstring theory that leads to the usual string theory as a
result of a certain spontaneous breakdown of higher spin
symmetries. Another motivation came from conjectured duality of a
conformal ${\cal N}=4$ SYM theory and a theory of type $IIB$
superstring in $AdS_5 \times S^5$ background \cite{malda}. Recent
discussion of this theme in the context of various limits in $AdS$
superstring may be found in \cite{ts1}. As is well known,
quantization of GS superstring propagating in flat space is
straightforward only in the light-cone gauge. It is the light-cone
gauge that removes unphysical degrees of freedom explicitly and
reduces the action to quadratical form in string coordinates. The
light-cone gauge in string theory implies the corresponding
light-cone formulation for target space fields. In the case of
strings in AdS background this suggests that we should study a
light-cone form dynamics of {\it target space fields} propagating
in AdS space-time. It is expected that AdS massive fields form
spectrum of states of AdS strings. Therefore understanding a
light-cone description of AdS massive target space fields might be
helpful in discussion of various aspects of AdS string dynamics.
This is what we are doing in this paper.

Let us first formulate the main problem we solve in this letter.
Fields propagating in $AdS_d$ space are associated with
positive-energy unitary lowest weight representations of
$SO(d-1,2)$ group. A positive-energy lowest weight irreducible
representation of the $SO(d-1,2)$ group denoted as $D(E_0,{\bf
h})$, is defined by $E_0$, the lowest eigenvalue of the energy
operator, and by ${\bf h}=(h_1,\ldots h_\nu)$,
$\nu=[\frac{d-1}{2}]$, which is the highest weight of the unitary
representation of the $SO(d-1)$ group. The highest weights $h_i$
are integers and half-integers for bosonic and fermionic fields
respectively. In this paper we investigate the fields whose $E_0$
and ${\bf h}$ are given by
\begin{eqnarray}\label{bose01} && E_0 > h_1 + d -3\,,\qquad {\bf h}= (h_1,0,\ldots, 0)
\\[7pt]
\label{fere01} && E_0 > h_1 + d -3\,,\qquad {\bf h}=\left\{
\begin{array}{ll}
(h_1,\frac{1}{2},\ldots,\frac{1}{2},\frac{1}{2}) \qquad & d-\hbox{
even }
\\[7pt]
(h_1,\frac{1}{2},\ldots,\frac{1}{2},\pm\frac{1}{2})& d-\hbox{
odd}\end{array}\right.
\end{eqnarray} The fields in \rf{bose01} are massive bosonic
fields while the ones in \rf{fere01} are fermionic fields. The
massive fields in \rf{bose01},\rf{fere01} with $h_1\geq 1$, are
referred to as totally symmetric fields\footnote{We note that the
case ${\bf h}=(1,0,\ldots,0)$ corresponds to spin one massive
field, the case ${\bf h}=(2,0,\ldots,0)$ is the massive spin two
field.}. In manifestly Lorentz covariant formulation the
bosonic(fermionic) totally symmetric massive representation is
described by a set of the tensor(tensor-spinor) fields whose
$SO(d-1,1)$ space-time tensor indices have the structure of the
respective Young tableauxes with one row. Covariant actions for
the {\it bosonic} totally symmetric massive fields in $AdS_d$
space were found in \cite{zin}.\footnote{Massive self-dual spin
fields in $AdS_3$ were investigated in \cite{rrm01088}. Spin two
massive fields were studied in \cite{pol}. Discussion of massive
totally symmetric fields in $(A)dS_d$, $d\geq 4$, may be found in
\cite{des}.}. {\it Fermionic} massive totally symmetry $AdS_d$
fields with arbitrary $E_0$ and ${\bf h}$ have not been described
at the field theoretical level so far \footnote{Group theoretical
description of various massive representation via oscillator
method may be found, e.g., in \cite{gunmin}. Lorentz covariant
equations of motion for $AdS_5$ self-dual massive fields with
special values of $E_0$ were discussed in \cite{sez2}. Light cone
actions for $AdS_5$ self-dual massive fields with arbitrary values
of $E_0$ were found in \cite{rrm02226}.}. In this paper we develop
a light-cone gauge formulation for such fields at the action
level. Using a new version of the old light-cone gauge formalism
in $AdS$ space developed in \cite{rrm99217}, we describe both the
bosonic and fermionic fields on an equal footing. Since, by
analogy with flat space, we expect that a quantization of the
Green-Schwarz $AdS$ superstring with Ramond-Ramond charge will be
straightforward only in the light-cone gauge \cite{mt3} it seems
that from the stringy perspective of $AdS/CFT$ correspondence the
light-cone approach is the fruitful direction to go.

\section{Light-cone gauge action and its global symmetries}

In this section we present new version of the old light cone
formalism developed in \cite{rrm99217}. Let $\phi(x)$ be a bosonic
arbitrary spin field propagating in $AdS_d$ space. If we collect
spin degrees of freedom in a ket-vector $|\phi\rangle$ then a
light-cone gauge action for $\phi$ can be cast into the following
`covariant form'\cite{rrm99217}\footnote{ We use parametrization
of $AdS_d$ space in which
$ds^2=(-dx_0^2+dx_i^2+dx_{d-1}^2+dz^2)/z^2$. Light-cone
coordinates in $\pm$ directions are defined as $x^\pm=(x^{d-1} \pm
x^0)/\sqrt{2}$ and $x^+$ is taken to be a light-cone time. We
adopt the conventions:
$\partial^i=\partial_i\equiv\partial/\partial x^i$,
$\partial_z\equiv\partial/\partial z$, $\partial^\pm=\partial_\mp
\equiv \partial/\partial x^\mp$, $z\equiv x^{d-2}$ and use indices
$i,j =1,\ldots, d-3$; $I,J=1,\ldots, d-2$. Vectors of $so(d-2)$
algebra are decomposed as $X^I=(X^i,X^z)$.}

\be\label{lcact} S_{l.c.} =\frac{1}{2}\int d^dx \langle
\phi^\dagger|\bigl(\Box -\frac{1}{z^2}A\bigr)|\phi\rangle\,,
\qquad \Box = 2\partial^+\partial^- + \partial_i^2+\partial_z^2\,.
\ee An operator $A$ does not depend on space-time coordinates and
their derivatives. This operator acts only on spin indices
collected in ket-vector $|\phi\rangle$. We call the operator $A$
the $AdS$ mass operator.

We turn now to discussion of global $so(d-1,2)$ symmetries of the
light-cone gauge action. The choice of the light-cone gauge spoils
the manifest global symmetries,  and in order to demonstrate that
these global invariances are still present one needs to find the
Noether charges which generate them\footnote{These charges play  a
crucial role  in formulating interaction vertices in field theory.
Application of Noether charges in formulating superstring field
theories may be found in \cite{gsb}}. Noether charges (or
generators) can be split into kinematical and dynamical
generators. For $x^+=0$ the kinematical generators are quadratic
in the physical field $|\phi\rangle$, while the dynamical
generators receive corrections in interaction theory.  In this
paper we deal with free fields. At a quadratic level both
kinematical and dynamical generators have the following standard
representation in terms of the physical light cone field
\cite{rrm99217} \be\label{hatG} \hat{G}=\int
dx^-d^{d-2}x\langle\partial^+\phi|G|\phi\rangle\,. \ee
Representation for the kinematical generators in terms  of
differential operators $G$ acting on the physical field
$|\phi\rangle$ is given by
\begin{eqnarray}
\label{pi}&& P^i=\partial^i\,, \qquad  P^+=\partial^+\,,
\\
&& D=x^+ P^- +x^-\partial^++x^I\partial^I+\frac{d-2}{2}\,,
\\
&& J^{+-}=x^+ P^- -x^-\partial^+\,,
\\
&& J^{+i}=x^+\partial^i-x^i\partial^+\,,
\\
&& J^{ij} = x^i\partial^j-x^j\partial^i + M^{ij}\,,
\\
&& K^+ = -\frac{1}{2}(2x^+x^-+x^Jx^J)\partial^+ + x^+D\,,
\\
&& K^i = -\frac{1}{2}(2x^+x^-+x^Jx^J)\partial^i +x^i
D+M^{iJ}x^J+M^{i-}x^+\,,
\end{eqnarray}
while a representation for the dynamical generators takes the form
\begin{eqnarray}
&& P^-=-\frac{\partial_J^2}{2\partial^+}
+\frac{1}{2z^2\partial^+}A\,,
\\
&& J^{-i}=x^-\partial^i-x^i P^-
+M^{-i}\,,\\
\label{km}&& K^-=-\frac{1}{2}(2x^+x^-+x_I^2) P^- +
x^-D+\frac{1}{\partial^+}x^I\partial^JM^{IJ}
-\frac{x^i}{2z\partial^+}[M^{zi},A] +\frac{1}{\partial^+}B\,,\ \ \
\ \
\end{eqnarray}
where $M^{-i}=-M^{i-}$ and \be M^{-i} \equiv
M^{iJ}\frac{\partial^J}{\partial^+}
-\frac{1}{2z\partial^+}[M^{zi},A]\,. \ee Operators $A$, $B$,
$M^{IJ}$ are acting only on spin degrees of freedom of wave
function $|\phi\rangle$. $M^{IJ}=M^{ij},M^{zi}$ are spin operators
of the $so(d-2)$ algebra \be\label{d2comrel}
[M^{IJ},M^{KL}]=\delta^{JK}M^{IL} +3 \hbox{ terms}\,,\ee while the
operators $A$ and $B$ admit the following representation
\begin{eqnarray}\label{adsope} A & = & \frac{1}{2}M^{IJ}M^{IJ}
+\frac{d(d-2)}{4}-\langle
Q_{AdS}\rangle +2B^z+ M^{zi}M^{zi}\,,
\\[6pt]
\label{bope} B & = & B^z + M^{zi}M^{zi}\,. \end{eqnarray} $\langle
Q_{AdS}\rangle$ is eigenvalue of the second order Casimir operator
of the $so(d-1,2)$ algebra for the representation labelled by
$D(E_0,{\bf h})$: \be\label{casope1} -\langle Q_{AdS}\rangle
=E_0(E_0+1-d)+ \sum_{\sigma=1}^\nu h_\sigma (h_\sigma -2\sigma
+d-1)\,,\ee while $B^z$ is $z$-component of $so(d-2)$ algebra
vector $B^I$ which satisfies the defining equation\footnote{We use
the notation $(M^3)^{[I|J]}\equiv \frac{1}{2}M^{IK}M^{KL}M^{LJ}
-(I\leftrightarrow J)$, $M^2\equiv M^{IJ}M^{IJ}$. Throughout this
paper we use a convention $N\equiv d-2$.} \be\label{basequ0}
[B^I,B^J] +(M^3)^{[I|J]} + (\langle Q_{AdS}\rangle -\frac{1}{2}M^2
-\frac{N(N-1)+2}{2})M^{IJ}\approx 0\,. \ee Here we use sign
$\approx$ to write instead of equations $X|\phi\rangle=0$
simplified equations $X\approx 0$. As was noted the operator $B^I$
transforms in vector representation of the $so(d-2)$ algebra
\be\label{bId2tra}
[B^I,M^{JK}]=\delta^{IJ}B^K-\delta^{IK}B^J\,.\ee Making use of the
formulas above-given one can check that the light-cone gauge
action \rf{lcact} is invariant with respect to the global
symmetries generated by $so(d-1,2)$ algebra taken to be in the
form $\delta_{\hat{G}} |\phi\rangle = G|\phi\rangle$. To summarize
a procedure of finding light cone description consists of the
following steps:

i) choose form of realization of spin degrees of freedom of the
field $|\phi\rangle$

ii) fix appropriate representation for the spin operator $M^{IJ}$;

iii) using formula \rf{casope1} evaluate an eigenvalue of the
Casimir operator;

iv) find solution to the defining equations for the operator $B^I$
\rf{basequ0}.

Now following this procedure we discuss bosonic and fermionic
fields in turn.

\section{Bosonic fields}

To discuss field theoretical description of massive AdS field we
could use $so(d-1)$ totally symmetric traceless tensor field
$\Phi^{\hat{I}_1\ldots \hat{I}_s}$, $\hat{I}=1',1,\ldots,d-2$.
Instead of this we prefer to decompose such field into traceless
totally symmetric tensors of $so(d-2)$ algebra $\phi^{I_1\ldots
I_{s'}}$, $I=1,\ldots, d-2$; $s'=0,1,\ldots, s$: \be
\Phi^{\hat{I}_1\ldots \hat{I}_s}=\sum_{s'=0}^s \oplus\,
\phi^{I_1\ldots I_{s'}}\,. \ee As usual to avoid cumbersome tensor
expressions we introduce creation and annihilation oscillators
$\alpha^I$ and $\bar{\alpha}^I$ \be\label{bososc}
[\bar{\alpha}^I,\,\alpha^J]=\delta^{IJ}\,,\qquad
\bar{\alpha}^I|0\rangle =0\,, \ee and make use of ket-vectors
$|\phi_{s'}\rangle$ defined by \be\label{genfun1}
|\phi_{s'}\rangle \equiv \alpha^{I_1}\ldots \alpha^{I_{s'}}
\phi^{I_1\ldots I_{s'}}(x)|0\rangle\,. \ee The $|\phi_{s'}\rangle$
satisfies the following algebraic constraints
\begin{eqnarray}
\label{homcon1}&& (\alpha\bar{\alpha}-s')|\phi_{s'} \rangle
=0\,,\qquad \alpha\bar{\alpha}\equiv \alpha^I\bar\alpha^I
\\
\label{tracon1}&& \bar\alpha^I\bar{\alpha}^I|\phi_{s'}
\rangle=0\,,\qquad \quad \hbox{tracelessness}\,.
\end{eqnarray}
Eq.\rf{homcon1} tells us that $|\phi_{s'}\rangle$ is a polynomial
of degree $s'$ in oscillator $\alpha^I$. Tracelessness of the
tensor fields $\phi^{I_1\ldots I_{s'}}$ is reflected in
\rf{tracon1}. The spin operator $M^{IJ}$ of the $so(d-2)$ algebra
for the above defined fields $|\phi_{s'}\rangle$ takes then the
form \be\label{spiopemij1} M^{IJ}
=\alpha^I\bar\alpha^J-\alpha^J\bar\alpha^I\,.\ee We are going to
connect our spin $s$ field with the unitary representation
labelled by $D(E_0,{\bf h})$ \rf{bose01} whose $h_1$ is identified
with spin value $s$:\footnote{This identification can be proved
rigourously by exploiting the procedure of Sec.6 in
Ref.\cite{rrm02226}.} \be\label{hside1} h_1= s\,.\ee Eigenvalue of
the second order Casimir operator for totally symmetric
representation under consideration is given then by (see
\rf{bose01},\rf{casope1},\rf{hside1}) \be\label{casope2} -\langle
Q_{AdS}\rangle =E_0(E_0+1-d) +s(s+d-3)\,.\ee An action of the
operator $B^I$ on the physical fields $|\phi_{s'}\rangle$ is found
to be \be\label{defbi} B^I|\phi_{s'}\rangle =a_{s'}A_{s'-1}^I
|\phi_{s'-1}\rangle +b_{s'}\bar{\alpha}^I|\phi_{s'+1}\rangle\,,
\ee where the coefficients $a_{s'}$, $b_{s'}$ depend on $E_0$, $s$
and $s'$ \be\label{abfinsol1} a_{s'} = b(E_0,s,s'-1)\,, \qquad
b_{s'} = b(E_0,s,s')\,,\ee and a function $b(E_0,s,s')$ is defined
to be \be b(E_0,s,s') \equiv
\left(\frac{(s-s')(s+s'+N-1)(E_0-s'-N)(E_0+s'-1)}{2s'+N}\right)^{1/2}\,.
\ee Operator $A_{s'}^I$ which appears in definition of the
operator $B^I$ \rf{defbi} is given by \be A_{s'}^I \equiv \alpha^I
-\frac{\alpha^2\bar{\alpha}^I}{2s'+N-2}\,,\qquad \alpha^2\equiv
\alpha^I\alpha^I\,. \ee

Let us outline procedure of derivation above-mentioned results.
Most difficult problem is to find solution to the defining
equation \rf{basequ0}. Fortunately, for the case of totally
symmetric fields this equation simplifies due to the following
relation for $M^{IJ}$: \be\label{bosm3}
(M^3)^{[I|J]}=(-\frac{1}{2}M^2+\frac{(N-2)(N-3)}{2})M^{IJ}\,. \ee
This formula can be checked directly by using representation for
$M^{IJ}$ given in \rf{spiopemij1}. Plugging  \rf{bosm3} in
\rf{basequ0} we get the following simplified form of defining
equation
 \be \label{adsbbm1} [B^I,B^J] +
(\langle Q_{AdS}\rangle -M^2 -2N +2)M^{IJ}\approx 0\,. \ee
Applying this equation to $|\phi_{s'}\rangle$ and using \rf{defbi}
we get a relationship for the coefficients $a_{s'}$ and $b_{s'}$
\be\label{absol1} a_{s'+1}b_{s'} =
\frac{(s-s')(s+s'+N-1)(E_0-s'-N)(E_0+s'-1)}{2s'+N}\,. \ee Now we
exploit a requirement the operator $B^I$ be hermitian\footnote{We
use anti-hermitian representation for generators of $so(d-1,2)$
algebra \rf{hatG}. This implies that the spin operator $M^{IJ}$
should be anti-hermitian, while the operators $A$ and $B^I$ should
be hermitian.} \be\label{bihercon} \langle\psi||B^I\phi\rangle =
\langle B^I\psi||\phi\rangle\,,\ee with respect to scalar product
defined by \be\label{norm1} \langle\psi||\phi\rangle\equiv
\sum_{s'=0}^s\langle\psi_{s'}||\phi_{s'}\rangle\,. \ee This
requirement gives the relation $a_{s'} =  {b}_{s'-1}^*$. Making
use of this relation in \rf{absol1} we arrive at the final
solution given in \rf{abfinsol1}. Some helpful formulas to
evaluate commutator $[B^I,B^J]$ are given by
\begin{eqnarray}
&& A_{s'}^I A_{s'-1}^J  -(I\leftrightarrow J) =0\,,
\\[6pt]
&& A_{s'}^I \bar{\alpha}^J - (I\leftrightarrow J) =M^{IJ}\,,
\\
&& \bar{\alpha}^I A_{s'}^J - (I\leftrightarrow J) =-
\frac{2s'+N}{2s'+N-2} M^{IJ}\,.
\end{eqnarray}

\section{Fermionic fields}

Light cone action for fermionic fields takes the following form
\be S_{l.c.}^{ferm} = \int d^dx \langle \psi^\dagger| \frac{\rm
i}{2\partial^+}(\Box - \frac{1}{z^2}A)|\psi\rangle\,. \ee This
action is invariant with respect to transformations $\delta |\psi
\rangle = G^{ferm}|\psi\rangle$, where differential operators
$G^{ferm}$ are obtainable from the ones for bosonic fields
(\rf{pi}-\rf{km}) by making there the following substitution $x^-
\rightarrow x^- +\frac{1}{2\partial^+}$. In addition to this in
expressions for generators in (\rf{pi}-\rf{km}) we should use the
spin operator $M^{IJ}$ suitable for fermionic fields. Defining
equation \rf{basequ0} for the operator $B^I$ does not change.
Before to discuss concrete form of the spin operators $M^{IJ}$ and
$B^I$ we should fix a field theoretical realization of spin
degrees of freedom collected in $|\psi\rangle$.

To discuss field theoretical description of massive AdS field we
could used totally symmetric traceless and $\gamma$-transversal
tensor-spinor field $\Psi^{\hat{I}_1\ldots \hat{I}_s\alpha}$,
$\hat{I}=1',1,\ldots,d-2$ which corresponds to irreducible spin
$s+\frac{1}{2}$ representation of $so(d-1)$ algebra. Instead of
this we prefer to decompose such field into traceless totally
symmetric tensor-spinor fields of $so(d-2)$ algebra
$\psi^{I_1\ldots I_{s'}\alpha}$, $I=1,\ldots, d-2$;
$s'=0,1,\ldots, s:$\footnote{Details of such a decomposition may
be found in Appendix B. The $\psi^{I_1\ldots I_{s'}\alpha}$ are
obtainable from $2^{[d/2]}$ Dirac tensor-spinor fields of
$so(d-1,1)$ algebra:
$\psi=\frac{1}{2}\gamma^-\gamma^+\psi_{Dirac}$. We use
$2^{[d/2]}\times 2^{[d/2]}$ -Dirac $\gamma$-matrices:
$\{\gamma^a,\gamma^b\}=2\eta^{ab}$, $\eta^{ab}=(-1,+1,\ldots,+1)$,
$a,b=0,1,\ldots, d-1$. In light cone frame we use a decomposition
$\gamma^a=\gamma^+,\gamma^-,\gamma^I$, where $\gamma^\pm\equiv
(\gamma^{d-1}\pm \gamma^0)/\sqrt{2}$. $I=1,\ldots, d-2$.} \be
\Psi^{\hat{I}_1\ldots \hat{I}_s \alpha}=\sum_{s'=0}^s \oplus\,
\psi^{I_1\ldots I_{s'}\alpha}\,. \ee As before to avoid cumbersome
expressions we exploit creation and annihilation oscillators
$\alpha^I$ and $\bar{\alpha}^I$ \rf{bososc} and make use of
ket-vectors $|\psi_{s'}\rangle$ defined by \be\label{genfun2}
|\psi_{s'}\rangle \equiv \alpha^{I_1}\ldots \alpha^{I_{s'}}
\psi^{I_1\ldots I_{s'}\alpha}(x)|0\rangle\,. \ee Here and below
spinor indices are implicit. The $|\psi_{s'}\rangle$ satisfies the
following algebraic constraints
\begin{eqnarray}
\label{homcon2}&& (\alpha\bar{\alpha}-s')|\psi_{s'} \rangle =0\,,
\\
\label{gamtra1}&& \gamma^I\bar\alpha^I|\psi_{s'} \rangle =0\,,
\quad \qquad\qquad \gamma- \hbox{transversality}
\\
\label{tracon2}&& \bar{\alpha}^I\bar\alpha^I|\psi_{s'}
\rangle=0\,,\qquad \qquad\quad \hbox{tracelessness}\,.
\end{eqnarray}
Eq. \rf{homcon2} tells us that $|\psi_{s'}\rangle$ is a polynomial
of degree $s'$ in oscillator $\alpha^I$. Tracelessness of the
tensor-spinor field $\psi^{I_1\ldots I_{s'}\alpha}$ is reflected
in \rf{tracon2}. Realization of spin operator $M^{IJ}$ on the
space of the ket-vectors $|\psi_{s'}\rangle$ is given by \be
M^{IJ} =M_b^{IJ}+\frac{1}{2}\gamma^{IJ}\,,\qquad M^{IJ}_b\equiv
\alpha^I\bar\alpha^J-\alpha^J\bar\alpha^I\,,\qquad
\gamma^{IJ}\equiv\frac{1}{2}(
\gamma^I\gamma^J-\gamma^J\gamma^I)\,.\ee We are going to connect
our massive tensor-spinor field with unitary representation
labelled by $D(E_0,{\bf h})$ \rf{bose01} whose $h_1$ is related
with integer $s$: \be\label{hside2} h_1= s+\frac{1}{2}\,.\ee
Eigenvalue of the second order Casimir operator for totally
symmetric fermionic representation is given then by (see
\rf{fere01},\rf{casope1},\rf{hside2}) \be\label{fercas1} -\langle
Q_{AdS}\rangle = E_0(E_0-1-N)
 +s(s+N)+\frac{N(N+1)}{8} \ee
An action of the operator $B^I$ on the physical  fermionic  fields
$|\psi_{s'}\rangle$ is found to be \be\label{defbi2} B^I
|\psi_{s'}\rangle = q_{s'}{\cal Y}_{s'}^I|\psi_{s'}\rangle +
a_{s'}{\cal A}_{s'-1}^I|\psi_{s'-1}\rangle +
b_{s'}\bar{\alpha}^I|\psi_{s'+1}\rangle\,.\ee As before the
coefficients $q_{s'}$, $a_{s'}$, $b_{s'}$ turn out to be functions
of $E_0$, $s$, $s'$ and are given by
\begin{eqnarray}
&& q_{s'} = \frac{2s+N}{2(2s'+N)}(E_0-\frac{N+1}{2})\,,
\\[5pt]
&& a_{s'} = f(E_0,s,s'-1)\,,
\\[5pt]
&& b_{s'} = f(E_0,s,s')\,, \end{eqnarray} where we use the
notation \be f(E_0,s,s')\equiv \left(
\frac{(s-s')(s+s'+N)(E_0-s'-N-\frac{1}{2})(E_0+s'-\frac{1}{2})}{2s'+N}\right)^{1/2}\,.
\ee Operators ${\cal A}_{s'}^I$ and ${\cal Y}_{s'}^I$ which enter
definition of basic operator $B^I$ \rf{defbi2} are given by \be
{\cal A}_{s'}^I \equiv \alpha^I
-\frac{\alpha^2\bar{\alpha}^I+(\gamma\alpha) \gamma^I}{2s'+N}\,,
\qquad  {\cal Y}_{s'}^I \equiv \gamma^I -
\frac{2(\gamma\alpha)\bar{\alpha}^I}{2s'+N-2}\,,\qquad
\gamma\alpha\equiv \gamma^I\alpha^I\,. \ee

Now let us outline procedure derivation of these results. We start
with general representation for $B^I$ given in \rf{defbi2} and the
problem is to find coefficients $q_{s'}$, $a_{s'}$, $b_{s'}$ which
satisfy the defining equation \rf{basequ0}. To this end we
evaluate expression for $(M^3)^{[I|J]}$
\begin{eqnarray}\label{m3}
(M^3)^{[I|J]}&\approx & M_b^{IJ}\Bigl(s'(s'+N-\frac{1}{2})
+\frac{2N(N-3)+5}{4}\Bigr)
\nonumber\\[7pt]
&+&\gamma^{IJ}\Bigl(\frac{1}{2}s' +\frac{N(N-3)+3}{8}\Bigr)
-\frac{2s'+ N-1}{4}(\gamma\alpha)\bar{S}^{IJ}\,,
\\[7pt]
\bar{S}^{IJ}&\equiv& \gamma^I\bar\alpha^J -
\gamma^J\bar\alpha^I\,,\nonumber
\end{eqnarray}
where a sign $\approx$ indicates that  \rf{m3} is valid by module
of terms which are equal to zero by applying to
$|\psi_{s'}\rangle$ i.e. to derive \rf{m3} we use constraints
\rf{homcon2}-\rf{tracon2}. After this using
Eqs.\rf{basequ0},\rf{fercas1},\rf{m3} we find solution for
$q_{s'}$ and product $a_{s'+1}b_{s'}$. Exploiting then the
hermicity condition for the operator $B^I$
\rf{bihercon},\rf{norm1} we arrive at the solution for $q_{s'}$,
$a_{s'}$, $b_{s'}$ given above. Some helpful formulas to evaluate
commutator $[B^I,B^J]$ are given by
\begin{eqnarray}
\label{aazer} && {\cal A}_{s'}^I {\cal A}_{s'-1}^J
-(I\leftrightarrow J) =0\,,
\\
&&[{\cal Y}_{s'}^I ,{\cal Y}_{s'}^J] =-\frac{4}{2s'+N-2}M_b^{IJ}
+2\gamma^{IJ} +4\frac{2s'+N-1}{(2s'+N-2)^2}
(\gamma\alpha)\bar{S}^{IJ}\,,\ \ \ \ \
\\
&&{\cal A}_{s'}^I \bar\alpha^J -(I\leftrightarrow J) =M_b^{IJ}
-\frac{(\gamma\alpha)\bar{S}^{IJ}}{2s'+N}\,,
\\
&&\bar\alpha^I{\cal A}_{s'}^J -(I\leftrightarrow J)
=-\frac{2s'+N+2}{2s'+N}M_b^{IJ} -\frac{2\gamma^{IJ}}{2s'+N}
+\frac{(\gamma\alpha)\bar{S}^{IJ}}{2s'+N}\,.
\end{eqnarray}

\section{Interrelation between lowest energy value $E_0$ and
mass parameter m}

In previous sections we have expressed our results in terms of
lowest eigenvalue of energy operator $E_0$. Because sometimes
formulation in terms of the standard mass parameter $m$ is
preferable we would like to derive interrelation between $E_0$ and
$m$. Before to going into details let of first present our
results.

Given massive fields with massive parameter $m$ corresponding to
unitary representation labelled by $D(E_0,{\bf h})$ we find the
following relationship between $E_0$ and $m$ (for even $AdS$
space-time dimension $d$; $\nu=(d-2)/2$)
\begin{eqnarray}
\label{bose0m}&& E_0 = \frac{d-1}{2} + \sqrt{ m^2 + \Bigl(h_k -k
+\frac{d-3}{2}\Bigr)^2\,}\,,\hspace{1.5cm} \hbox{ for bosonic
fields};
\\[8pt]
\label{fere0m}&& E_0 = m + h_k -k -2 +d\,,\,\,\, \hspace{4cm}
\hbox{ for fermionic fields}\,,
\end{eqnarray}
where a number $k$ is defined from the relation
\begin{equation}\label{mineq2}
h^{}_1=\ldots =h_k^{} > h_{k+1}^{} \ge h^{}_{k+2}\ge \ldots \ge
h_\nu^{}\geq 0.
\end{equation}
We remind that for bosonic fields the labels $h_\sigma$ are
integers while for fermionic fields the $h_\sigma$ are
half-integers. We note that relations \rf{bose0m},\rf{fere0m} are
valid also for those massive fields in odd dimensional $AdS_d$
whose $h_{(d-1)/2}=0$.

Now let us outline procedure of derivation these results. Let
$\phi_{m=0}^{\mu_1\ldots}$ be massless field in $AdS_d$. As was
demonstrated in \cite{rrmsit1},\cite{rrm98231} the massless fields
associated with unitary representation labelled by
$D(E_0^{m=0},{\bf h})$ should satisfy the equation of motion
\begin{equation}\label{eqmot4}
({\cal D}^2-E_0^{m=0}(E_0^{m=0}+1-d)+\sum_{\sigma=1}^\nu
h_\sigma)\phi_{m=0}^{\mu_1\ldots}=0\,,
\end{equation}
where ${\cal D}^2$ is a covariant D'Alembertian operator in
$AdS_d$ and $E_0^{m=0}$ is given by \be\label{e0m0}  E^{m=0}_0 =
h_k-k-2+d\,.\ee Eq. \rf{eqmot4} reflects the well-known fact that
equations of motion for $AdS$ massless field involve mass-like
term which is expressible in terms of $E_0^{m=0}$. As is well
known to discuss gauge invariant description of massive fields one
introduces the set of fields including some fields
$\phi_m^{\mu_1\ldots }$ which we shall refer to as leading field
plus Goldstone fields (sometimes referred to as Stueckelberg
fields). By definition, the structure of Lorentz indices of the
leading massive field $\phi_m^{\mu_1\ldots }$ is the same as the
one for massless field $\phi_{m=0}^{\mu_1\ldots }$. If we impose
on the leading field $\phi_m^{\mu_1\ldots }$ and the Goldstone
fields an appropriate covariant Lorentz gauge and tracelessnes
conditions then for the leading field one gets the equation
\begin{equation}\label{eqmot41}
({\cal D}^2- m^2-E_0^{m=0}(E_0^{m=0}+1-d) +\sum_{\sigma=1}^\nu
h_\sigma)\phi_m^{\mu_1\ldots}=0\,.
\end{equation}
One other hand an analysis of Ref.\cite{rrmsit1},\cite{rrm98231}
implies that the leading field satisfies the following equations
of motion
\begin{equation}\label{eqmot42}
({\cal D}^2- E_0(E_0+1-d) +\sum_{\sigma=1}^\nu
h_\sigma)\phi_m^{\mu_1\ldots}=0\,.
\end{equation}
Comparison of Eqs. \rf{eqmot41} and \rf{eqmot42} gives a
relationship  between $E_0$ and $m$ \be m^2 = E_0(E_0+1-d)-
E_0^{m=0}(E_0^{m=0}+1-d)\,.\ee Solution to this equation
corresponding to positive values of $E_0$ is given in \rf{bose0m}.

The same arguments can be applied to the fermionic fields. In this
case equation for massless field in $AdS_d$ is given by (see the
second Refs. in \cite{rrmsit1}) \be \label{eqmot6r} (\gamma^a
e_a^\mu D_{\mu L}+E_0^{m=0}+\frac{1-d}{2})\psi_{m=0}^{\alpha\,
\mu_1 \ldots } =0\,, \ee where $D_{\mu L}$ is a covariant
derivative with respect to local Lorentz rotation, $e_a^\mu$ is an
inverse of the vielbein $e_\mu^a$ and $E_0^{m=0}$ is given in
\rf{e0m0}. Because on the one hand the leading massive
tensor-spinor field $\psi_m^{\alpha\, \mu_1 \ldots }$ taken to be
in appropriate covariant Lorentz gauge satisfies an equation \be
\label{eqmot61r} (\gamma^a e_a^\mu D_{\mu
L}+m+E_0^{m=0}+\frac{1-d}{2})\psi_m^{\alpha\, \mu_1 \ldots } =0\,,
\ee and on the other hand this equation should be representable as
(see \cite{rrmsit1}) \be \label{eqmot62r} (\gamma^a e_a^\mu D_{\mu
L}+E_0+\frac{1-d}{2})\psi_m^{\alpha\, \mu_1 \ldots } =0\,, \ee we
find the relation for $m$ \be m = E_0 - E_0^{m=0}\,,\ee which
together with \rf{e0m0} leads to \rf{fere0m}.

In the AdS/CFT correspondence the $E_0$ is connected with
dimension of conformal operator as $E_0=\Delta$. The $\Delta$ for
bosonic massive totally antisymmetric and bosonic massive
symmetric spin two fields were evaluated in
Refs.\cite{l'Yi:1998eu},\cite{pol}. Our results coincides with the
ones obtained in these references. For instance for the case of
bosonic massive totally symmetric fields we have $h_1=s$, $k=1$
and this leads to \be E_0 =\Delta =\frac{d-1}{2} +\sqrt{m^2
+\Bigl(s+\frac{d-5}{2}\Bigr)^2\,}\,.\ee {}For the case of $s=2$
this is result of Ref.\cite{pol}.

{}For fermionic massive totally symmetric fields we have
$h_1=s+\frac{1}{2}$, $k=1$ and formula \rf{fere0m} leads to \be
E_0 = \Delta = m + s +d -\frac{5}{2}\,.\ee {}For particular value
of $s=1$ (Rarita-Schwinger field) appropriate $\Delta$ was
evaluated in Refs.\cite{Volovich:1998tj},\cite{Koshelev:1998tu}.
Note that our result taken to be for $s=1$ differs from the one
obtained in these references. The reason for this is that in this
paper we use normalization of mass parameter such that the point
$m=0$ corresponds to massless fields (see Eqs.
\rf{eqmot6r},\rf{eqmot61r}). In
Refs.\cite{Volovich:1998tj},\cite{Koshelev:1998tu} another
normalization was used.

Because while derivation of our results for $E_0$ we used
arguments based on gauge invariant formulation for massive fields
our relations \rf{bose0m},\rf{fere0m} are not applicable to the
scalar field and spin one-half field. Conformal dimension for
these fields are well known (see
\cite{Witten:1998qj},\cite{Henningson:1998cd}).

{\bf Conclusions}. The results presented here should have a number
of interesting applications and generalizations, some of which
are: i) In this paper we develop light cone formulation for
massive totally symmetric  fields. It would be interesting to
extend such formulation to the study of massive mixed symmetry
fields (see
Refs.\cite{brimet},\cite{alk164},\cite{deMedeiros:2003px}) and
then to apply  such formulation to the study of AdS/CFT
correspondence along the line of
Ref.\cite{Dobrev:2002xk},\cite{Bianchi:2003wx}. ii) As is well
known massive and massless fields can be connected via procedure
of dimensional reduction. Procedure of dimensional reduction
$AdS_{d}\rightarrow AdS_{d-1}$ was developed in \cite{rrm01088}
(for discussion of alternative reductions see
\cite{Biswas:2002nk},\cite{Bekaert:2003uc}). It would be
interesting to find mass spectra of AdS massive modes upon
dimensional reduction of massless AdS fields.

{\bf Acknowledgments}. This work was supported by the INTAS
project 03-51-6346, by the RFBR Grant No.02-02-17067, and RFBR
Grant for Leading Scientific Schools, Grant No. 1578-2003-2.

\setcounter{section}{0} \setcounter{subsection}{0}

\appendix{Interrelation between new and old light cone formulations}

In this appendix we explain interrelation of basic defining
equation for the operator $B^I$ and representation for the
operators $A$ and $B$ \rf{adsope},\rf{bope} with the old defining
equations given in Ref.\cite{rrm99217}. Defining equations for the
operators $A$ and $B$  take the form\cite{rrm99217}
\begin{eqnarray}
\label{defcon1r} && 2\{M^{zi},A\}-[[M^{zi},A],A]=0\,,
\\[6pt]
\label{defcon2r} && [M^{zi},[M^{zj},A]]+\{M^{iL},M^{Lj}\}
=-2\delta^{ij}B\,,
\\[6pt]
\label{casequ}&&A=2B+\frac{1}{2}M_{ij}^2+\frac{d(d-2)}{4}-\langle
Q_{AdS}\rangle\,,
\\[6pt]
\label{invequ}&&{} [A,M^{ij}]=0\,.
\end{eqnarray}
We note that Eq.\rf{invequ} tells us that the operators $A$ is
invariant with respect to $so(d-3)$ rotations. Eq.\rf{casequ} is
because of the second order Casimir operator of $so(d-1,2)$
algebra is diagonal in irreps labelled by $D(E_0,{\bf h})$.
Eqs.\rf{defcon1r},\rf{defcon2r} are consequences of commutators of
the $so(d-1,2)$ algebra. Introducing new operator $\tilde{B}$ by
relation \be B = \tilde{B} + M^{zi}M^{zi}\,, \ee and plugging the
representation for the operator $A$ given in \rf{casequ} in Eq.
\rf{defcon2r} we cast the above given system of equations into the
following form
\begin{eqnarray} \label{defcon1rr} &&{}
[[M^{zi},\tilde{B}],\tilde{B}] +(M^3)^{[z|i]} + (\langle
Q_{AdS}\rangle -\frac{1}{2}M^2 -\frac{d^2-5d+8}{2})M^{zi}=0\,,
\\[6pt]
\label{defcon2rr} &&
[M^{zi},[M^{zj},\tilde{B}]]+\delta^{ij}\tilde{B}=0\,,
\\[6pt]
&& A=\frac{1}{2}M_{IJ}^2+\frac{d(d-2)}{4}-\langle Q_{AdS}\rangle
+2\tilde{B}+M^{zi}M^{zi}\,,
\\[6pt]
\label{btinvcon}&& [\tilde{B},M^{ij}]=0\,.
\end{eqnarray}
While deriving of these formulas we use commutation relations of
$so(d-2)$ algebra spin operators $M^{IJ}=M^{ij},M^{zi}$ given in
\rf{d2comrel} and exploit the relation \be M^{IJ}M^{IJ}
=M^{ij}M^{ij}+ 2M^{zi}M^{zi}\,.\ee Our basic observation is that
if we introduce the quantities \be\label{bIdef} B^z \equiv
\tilde{B}\,, \qquad\qquad B^i\equiv [\tilde{B},M^{zi}]\,, \ee then
the operator $B^I=B^i,B^z$ transform as a vector under rotation
generated by $so(d-2)$ algebra. Indeed the second relation in
\rf{bIdef} and Eqs.\rf{defcon2rr},\rf{btinvcon} can be rewritten
as \be [M^{zi},B^j]=\delta^{ij}B^z\,, \qquad
[M^{zi},B^z]=-B^i\,,\qquad [M^{ij},B^z]=0\,.\ee These commutation
relations together with the ones following from $so(d-3)$
covariance \be [M^{ij},B^k]= \delta^{jk}B^i -\delta^{ik}B^j\,,\ee
imply that the operator $B^I$ is indeed transformed in vector
representation of $so(d-2)$ algebra (cf. \rf{bId2tra}). All that
remains is to analyse Eq.\rf{defcon1rr}. Making use of \rf{bIdef}
we can rewrite \rf{defcon1rr} as \be\label{bzbicon1} [B^z,B^i]
+(M^3)^{[z|i]} + (\langle Q_{AdS}\rangle -\frac{1}{2}M^2
-\frac{d^2-5d+8}{2})M^{zi}=0\,.\ee Now taking into account that a
state obtainable by acting the spin operator $M^{zj}$ on wave
function $|\phi\rangle$ should also belong to $|\phi\rangle$ we
conclude that constraint \rf{bzbicon1} should commute with
$M^{zj}$. This gives a new constraint \be\label{bzbicon2}
[B^i,B^j] +(M^3)^{[i|j]} + (\langle Q_{AdS}\rangle -\frac{1}{2}M^2
-\frac{d^2-5d+8}{2})M^{ij}=0\,.\ee Constraints
\rf{bzbicon1},\rf{bzbicon2} are collected into the ones given in
\rf{basequ0}. Thus we proved that the old light cone formalism of
Ref.\cite{rrm99217} is equivalent to the one used in this paper.
Remarkable features of new formalism are i) appearance of
$so(d-2)$ vector $B^I$ which was hidden in old formalism; ii)
manifest $so(d-2)$ invariance of defining equations for $B^I$.

\appendix{Decompositions of tensor-spinor fields}

Here we wish to describe an decomposition of totally symmetric
tensor-spinor field $\Psi^{\hat{I}_1\ldots \hat{I}_s\alpha}$
transforming in irreps of the $so(d-1)$ algebra in terms of
tensor-spinor fields $|\psi_{s'}\rangle$ which are irreps of the
$so(d-2)$ algebra. Consider a generating function \be
|\Psi\rangle\equiv \Psi^{\hat{I}_1\ldots
\hat{I}_s\alpha}\alpha^{\hat{I}_1}\ldots
\alpha^{\hat{I}_s}|0\rangle\,,\qquad
\alpha^{\hat{I}}=(\alpha^{1'},\alpha^I)\,, \ee which satisfies the
constraint \be \label{trco}
\alpha^{\hat{I}}\bar\alpha^{\hat{I}}|\Psi\rangle=s|\Psi\rangle\,,
\qquad
\bar{\alpha}^{\hat{I}}\bar\alpha^{\hat{I}}|\Psi\rangle=0\,,\qquad
\Gamma^{\hat{I}}\bar\alpha^{\hat{I}}|\Psi\rangle=0\,, \ee where to
define $\Gamma$-transversality constraint we use
$\Gamma^{\hat{I}}$-symbols defined by relations
\begin{eqnarray}
\Gamma^I =\gamma^I\,,\qquad \Gamma^{1'}=\left\{\begin{array}{ll}
{\rm i}^{(d-2)/2}\gamma^1\ldots \gamma^{d-2}\,, \qquad\qquad &
d-\hbox{ even};
\\[5pt]
{\rm i} & d-\hbox{ odd}.
\end{array}\right.
\end{eqnarray}
We start with analysis of the second constraint in (\ref{trco})
which can be considered  as the second order differential equation
with respect to oscillator variable $\alpha^{1'}$ \be
(\bar\alpha^{1'}\bar\alpha^{1'}+\omega^2)|\Psi(\alpha^{1'},\alpha^I)\rangle=0\,,
\qquad \omega^2=\bar{\alpha}^I\bar{\alpha}^I\,. \ee Obvious
solution to this equation is found to be \be\label{pp1p2}
|\Psi(\alpha^{1'},\alpha^I)\rangle =\cos (\omega
\alpha^{1'})|\Psi_s(\alpha^I)\rangle +\frac{\sin (\omega
\alpha^{1'})}{\omega}|\Psi_{s-1}(\alpha^I)\rangle\,, \ee where
$|\Psi_s\rangle$ and $|\Psi_{s-1}\rangle$ are rank $s$ and $s-1$
traceful tensor-spinors, i.e. they are reducible representations
of the $so(d-2)$ algebra. The solution (\ref{pp1p2}) reflects well
know fact that symmetric traceless rank $s$  tensor of $so(d-1)$
algebra can be decomposed into symmetric rank $s$ and $s-1$
traceful tensors of $so(d-2)$ algebra. The
$|\Psi_s\rangle$,$|\Psi_{s-1}\rangle$ satisfy the constraints $
(\alpha^I\bar{\alpha}^I-s)|\Psi_s\rangle=0$, $
(\alpha^I\bar\alpha^I-s+1)|\Psi_{s-1}\rangle=0$. Now plugging
\rf{pp1p2} into the third constraint in \rf{trco} we find a
relation \be\label{psipsi12} |\Psi_{s-1} \rangle =
\bar\alpha^I\gamma^I\Gamma^{1'}|\Psi_s\rangle\\,. \ee Taking into
account \rf{psipsi12} and \rf{pp1p2} we get a relation \be |\Psi
\rangle
=\exp(\alpha^{1'}\bar\alpha^I\gamma^I\Gamma^{1'})|\Psi_s\rangle\,,
\ee which tells us that traceless and $\Gamma$-transversal
tensor-spinor field $|\Psi\rangle$ (see constraints \rf{trco}) is
expressible in terms of one tracefull tensor-spinor
$|\Psi_s\rangle$. This tracefull tensor-spinor field in turn can
be decomposed into traceless tensor-spinors fields of $so(d-2)$
algebra $|\tilde\psi_{s'}\rangle$ \be |\Psi_s \rangle = \sum_{s'}
\oplus |\tilde\psi_{s'}\rangle\,,\qquad\qquad s'=s,s-2,s-4, \ldots
,s-2[s/2]\,. \ee The traceless tensor-spinor fields of $so(d-2)$
algebra $|\tilde\psi_{s'}\rangle$ satisfy by definition the
constraints \be \bar\alpha^I\bar\alpha^I
|\tilde\psi_{s'}\rangle=0\,, \qquad
(\alpha\bar\alpha-s')|\tilde\psi_{s'}\rangle=0\,. \ee Because the
tensor-spinor $|\tilde\psi_{s'}\rangle$ does not satisfy
$\gamma$-transversality constraint this field is a reducible
representation of the $so(d-2)$ algebra. It can be decomposed into
two irreducible representations of the $so(d-2)$ algebra by
formulas \be |\psi_{s'}\rangle =
(1-\frac{(\gamma\alpha)(\gamma\bar{\alpha})}{2s'+N-2}\Bigr)|\tilde\psi_{s'}\rangle\,,
\qquad \qquad |\psi_{s'-1}\rangle =
\gamma\bar\alpha|\tilde\psi_{s'}\rangle\,. \ee It is the set of
traceless and $\gamma$-transversal tensor-spinor fields
$|\psi_{s'}\rangle$ that we used to formulate light cone action
for massive fermionic representations of $so(d-1,2)$ algebra.

\end{document}